\documentclass{iau} 
\usepackage{graphicx}

\title[Secular Evolution in TIMER] 
{Kinematical Signatures of Disc Instabilities and Secular Evolution in the MUSE\\TIMER Survey}

\author[Dimitri A. Gadotti \& the TIMER Team]   
{Dimitri A. Gadotti$^1$, 
Adrian Bittner$^1$, 
Jesus Falc\'on-Barroso$^2$, 
Jairo M\'endez-Abreu$^2$
\and the TIMER Team$^3$}

\affiliation{$^1$European Southern Observatory\\[\affilskip]
$^2$ Instituto de Astrof\'isica de Canarias\\[\affilskip]
$^3$https://www.muse-timer.org/team}

\pubyear{2019}
\volume{353}  
\jname{Galactic Dynamics in the Era of Large Surveys}
\editors{M. Valluri, \& J. A. Sellwood  }
\begin{document}

\maketitle

\begin{abstract}
The MUSE TIMER Survey has obtained high signal and high spatial resolution integral-field spectroscopy data of the inner $\sim6\times6$\,kpc of 21 nearby massive disc galaxies. This allows studies of the stellar kinematics of the central regions of massive disc galaxies that are unprecedented in spatial resolution. We confirm previous predictions from numerical and hydrodynamical simulations of the effects of bars and inner bars on stellar and gaseous kinematics, and also identify box/peanuts via kinematical signatures in mildly and moderately inclined galaxies, including a box/peanut in a face-on inner bar. In 20/21 galaxies we find inner discs and show that their properties are fully consistent with the bar-driven secular evolution picture for their formation. In addition, we show that these inner discs have, in the region where they dominate, larger rotational support than the main galaxy disc, and discuss how their stellar population properties can be used to estimate when in cosmic history the main bar formed. Our results are compared with photometric studies in the context of the nature of galaxy bulges and we show that inner discs are identified in image decompositions as photometric bulges with exponential profiles (i.e., S\'ersic indices near unity).
\keywords{galaxies: bulges, galaxies: evolution, galaxies: kinematics and dynamics, galaxies: structure}
\end{abstract}

\firstsection 
\section{Introduction: the TIMER project}

The TIMER project is a survey with MUSE of 24 nearby barred galaxies with nuclear structures such as nuclear rings, inner discs, inner bars and nuclear spiral arms (see \cite{Dimitri}). The MUSE datacubes cover the central $1'\times1'$ providing 90\,000 high signal spectra per galaxy, with a spatial sampling of $0.2''$ (see Fig.\,1). One of the main goals of the project is to derive the star formation histories of the nuclear structures, which allows us to estimate the time of bar formation. This, in turn, is an indication of the time in which the main stellar disc settles.

We have offered a proof of concept in a pilot study, in which we show that the bar of the massive Virgo S0 galaxy NGC\,4371 is about 10\,Gyr old, with a formation redshift in the range $1.4<z<2.3$ (\cite{T1}). This result attests to the robustness of bars and suggests that discs settle first, and bars form earlier, in more massive galaxies. With the TIMER data, we will also test this downsizing picture, as the sample covers a mass range from $\approx10^{10}$\,M$_\odot$ to $\approx10^{11}$\,M$_\odot$.

\begin{figure}[t]
\begin{center}
 \includegraphics[width=0.6\columnwidth]{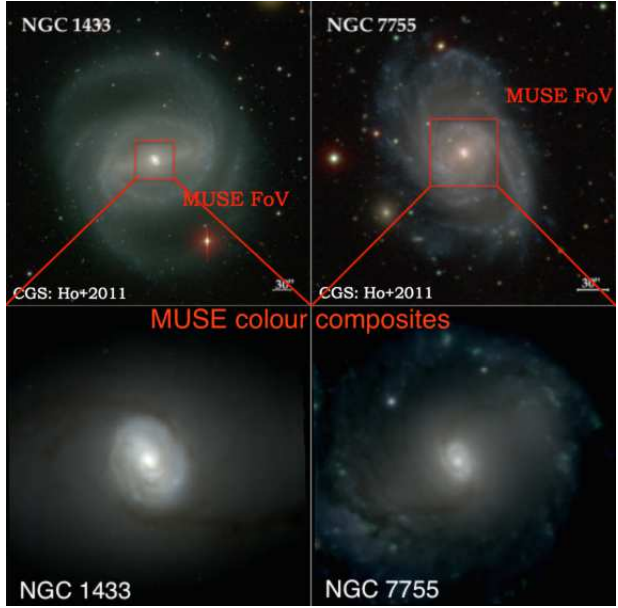} 
 \caption{Colour composites of two TIMER galaxies (NGC\,1433 and NGC\,7755) from the Carnegie Galaxy Survey (\cite{Ho}, top) and from our MUSE datacubes (bottom). The inner discs are clearly visible in the latter, as well as the dust lanes along the leading sides of the bar, feeding the inner disc with gas.}
   \label{fig1}
\end{center}
\end{figure}

\section{Inner discs built by bars}

\begin{figure}[t]
\begin{center}
 \includegraphics[width=1\columnwidth]{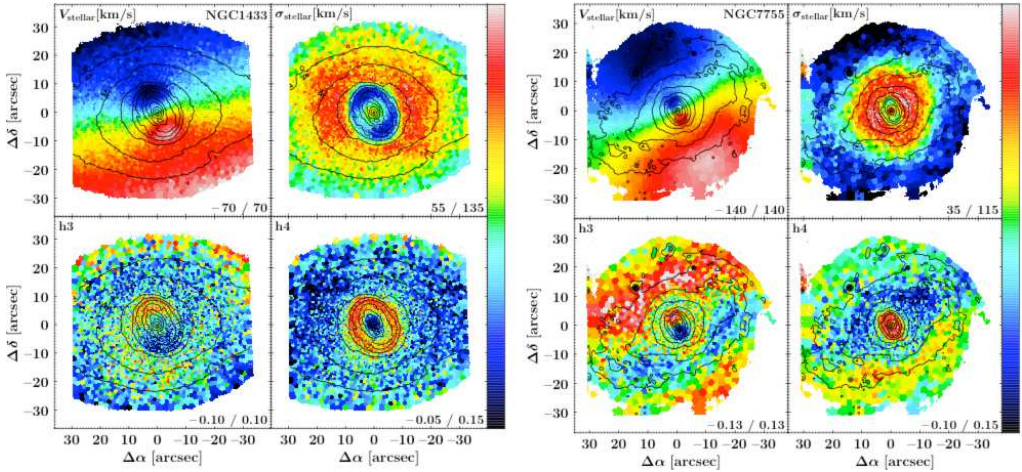} 
 \caption{Kinematical maps for NGC\,1433 and NGC\,7755, showing $v$, $\sigma$, h$_3$ and h$_4$, as indicated.}
   \label{fig2}
\end{center}
\end{figure}

\begin{figure}[t]
\begin{center}
 \includegraphics[width=0.55\columnwidth]{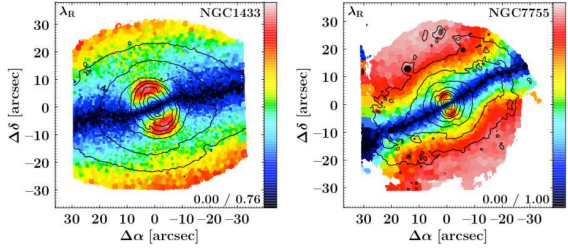} 
 \caption{Maps of $\lambda$ for NGC\,1433 and NGC\,7755.}
   \label{fig3}
\end{center}
\end{figure}

In Fig.\,2 we show the kinematical maps derived with the GIST pipeline (\cite{Adrian}) for the two TIMER galaxies shown in Fig.\,1. For 20 of the 21 galaxies studied so far we find the same results: the nuclear structures can be characterised by an inner disc that rotates fast, has low velocity dispersion $\sigma$, and shows an anti-correlation between the velocity $v$ and the high-order moment of the line-of-sight velocity distribution h$_3$ (a signature of near-circular orbits). In addition, it shows elevated values of another high-order moment, h$_4$, which indicates that the inner disc is an additional component to the main galaxy disc, not only the inner part of the main disc.

This conclusion is strengthened by the results shown in Fig.\,3. This figure shows maps of $\lambda$ (as defined in \cite{Eric} but for each spatial bin), which indicates the amount of dynamical support provided by ordered motion. At the radii where they reside, inner discs have more rotational support than the main galaxy disc. These properties are consistent with the picture in which bars bring gas to the central region, thus building inner stellar discs (see, e.g., \cite{Lia}).

The h$_3$ maps in Fig.\,3 also show that in the region where the bar dominates, instead of finding an anti-correlation between $v$ and h$_3$, one finds a {\em correlation}, which indicates elongated orbits. This agrees with the expectation from models of barred galaxies, such as those presented in \cite{Iannuzzi}.

\section{Box/peanuts}

\begin{figure}[t]
\begin{center}
 \includegraphics[width=0.5\columnwidth]{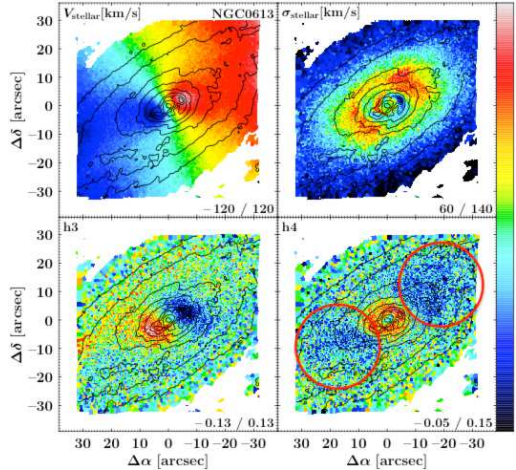} 
 \caption{Kinematical maps for NGC\,613. The h$_4$ map (bottom right) clearly shows two minima along the bar (highlighted by red circles), indicating the presence of a box/peanut.}
   \label{fig4}
\end{center}
\end{figure}

We also find in several galaxies signatures of the presence of box/peanuts (the inner part of bars that evolves vertically, away from the plane of the main disc), which are difficult to see in galaxies that are not edge-on. Box/peanuts induce h$_4$ minima along the bar (\cite{Victor}; \cite{Jairo1}) as can be seen, e.g., in NGC\,613 (see Fig. 4). We also detected for the first time the presence of a box/peanut in an inner bar, (in NGC\,1291; \cite{Jairo2}), which demonstrates that the dynamics of inner and main bars is similar (see also \cite{Adriana}).

\section{Exponential bulges are inner discs}

\begin{figure}[t]
\begin{center}
 \includegraphics[width=0.6\columnwidth]{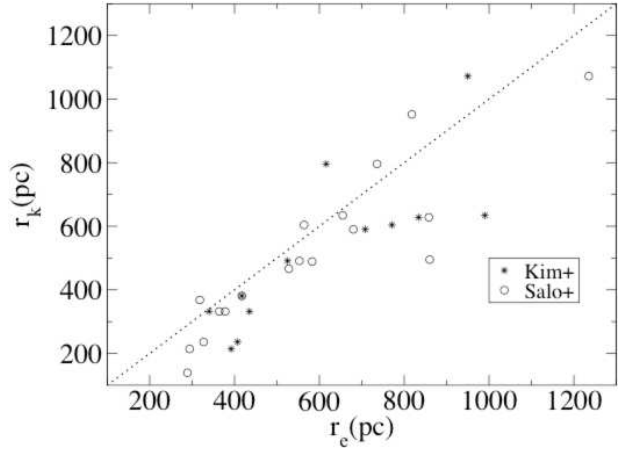} 
 \caption{Correlation between the effective radii (r$_{\rm e}$) of central exponential components found through multi-component image decompositions and r$_{\rm k}$, the sizes of the inner discs as derived with the TIMER kinematical maps. The latter are defined as the radius of the peak in $\lambda$ within the inner disc. This match shows that the so-called pseudo-bulges or disc-like bulges are in fact rapidly rotating inner discs.}
   \label{fig5}
\end{center}
\end{figure}

Finally, we investigate the photometrical properties of the inner discs identified kinematically, by examining the results from two independent studies that employ multi-component image decompositions (\cite{Tina}; \cite{Salo}). We find that, typically, the luminosity radial profile of our inner discs is better described by a S\'ersic function with index near unity, i.e., they have near exponential profiles, as do main galaxy discs.

Furthermore, in Fig.\,5 we compare the effective radius of the photometric components derived through image decomposition with the size of the inner discs as derived from the TIMER kinematical maps. The latter are defined as the radius of the peak in $\lambda$ within the inner disc. One sees that both measurements are strongly correlated.

Taken altogether, these results show that the central exponential components found through photometric analyses (often also called pseudo-bulges or disc-like bulges) are in fact inner discs built through bar-driven secular evolution processes. A full-fledged article reporting this work will appear shortly in D.A. Gadotti et al. (in prep.).

\end{document}